\documentclass{JHEP3}
\newcommand{\beq}{\begin{equation}}
\newcommand{\eeq}{\end{equation}}
\newcommand{\phib}{\ensuremath{\overline{\phi}}}

\newcommand{\psib}{\ensuremath{\overline{\psi}}}

\newcommand{\KD}{K\"{a}hler-Dirac }
\newcommand{\cF}{\ensuremath{{\cal F}}}

\usepackage{epsfig}

\title{On the restoration of supersymmetry in twisted two-dimensional
lattice Yang-Mills theory}

\author{Simon Catterall\\
        Department of Physics, Syracuse University, Syracuse, NY 13244, USA\\
        E-mail: \email{smc@physics.syr.edu}\\
        }

\preprint{SU-4252-840}  

\abstract{We study a discretization of ${\cal N}=2$ super Yang-Mills theory
which possesses a single exact 
supersymmetry at non-zero lattice spacing. This supersymmetry
arises after a reformulation of the theory in terms of {\it twisted}
fields. In this paper we derive the action
of the other twisted supersymmetries on the component
fields and study, using Monte Carlo simulation, a series of 
corresponding Ward identities. Our 
results for $SU(2)$ and $SU(3)$ 
support a restoration of these additional supersymmetries 
without fine tuning in the infinite volume continuum limit. 
Additionally we present
evidence supporting 
a restoration of (twisted) rotational invariance in the same
limit. Finally we have
examined the distribution of scalar field eigenvalues
and find evidence for power law tails extending out to large eigenvalue.
We argue that these tails indicate that the classical moduli
space does not survive in the quantum theory.
}


\begin{document} 
\section{Introduction}
Supersymmetric theories have long been of interest to particle
physicists both from a phenomenological and theoretical
perspective. However for many years the study of such systems
on the lattice was problematic. Supersymmetry is typically broken
at the classical level in such theories and this makes it
difficult if not impossible to construct supersymmetric continuum
limits for such theories -- see, for example the
reviews \cite{plenary,kplenary,fplenary} and references therein.

Recently we have developed a formalism for discretizing supersymmetric
theories in $D$ dimensions with ${\cal N}=p\times 2^D,p=1,2,\ldots$
supercharges in which one or more supersymmetries are preserved
exactly at non-zero lattice spacing.\footnote{Strictly the construction
yields a lattice theory with complexified lattice fields but 
theoretical
arguments and numerical work have provided evidence that the supersymmetry
is retained at the quantum level when the fields are restricted to
be real.}
The discretization proceeds from a 
reformulation of the continuum theory in terms of {\it twisted} fields.
Such a procedure naturally exposes a nilpotent supercharge $Q$ and
generically leads to an action which can be written in $Q$-exact form.
These properties 
allow us to construct supersymmetric lattice actions which preserve
the corresponding supersymmetry \cite{top}.
Several lattice models have been proposed starting from
these twisted continuum formulations \cite{sugino,noboru,tomo}. In this paper
we will investigate the discretization developed in
\cite{n=2} and \cite{n=4}.
In this case the resulting lattice theories yield 
an alternative geometrical reformulation of the orbifold
constructions of Kaplan et al. \cite{kaplan}. The explicit
connection between the the twisted and orbifold lattice
constructions was
shown recently by Unsal \cite{unsal}. The two
constructions share the same naive
continuum
limit and preserve the same number of supersymmetries. Indeed, we show
later that linear combinations of the scalar and rank 2 tensor
supercharges occurring in the twisted
formulation of ${\cal N}=2$ SYM yield, in the naive
continuum limit,
the conserved nilpotent charge of the equivalent orbifold
model. Thus the two approaches to lattice supersymmetry,
while differing in their details, depend on essentially the same
underlying mechanisms for preserving supersymmetry at finite lattice
spacing. 
 
The fermion fields that appear in the twisted
formulations
can be naturally embedded as
components of one or more K\"{a}hler-Dirac fields.
This geometrical interpretation of the fermions makes it
then almost trivial to discretize the theory in a way
which avoids
the notorious problem of fermion doubling. 
Furthermore, using the discretization
prescription developed in \cite{n=2}, the model may then be 
studied using numerical simulation
techniques \cite{susysim}.

In this paper we extend previous work by deriving the action of
the remaining supersymmetries on the twisted fields and use 
Monte Carlo simulation to examine a series of Ward identities which
yield information on the restoration of full supersymmetry in the
continuum limit. The next two sections discuss the continuum twisted
theory and its transcription to the lattice. We then derive the
remaining twisted supersymmetries and go on to describe our
numerical results concerning both the Ward identities and
the question of the restoration of rotational invariance. The
bulk of our simulations are conducted in the {\it phase quenched
approximation}
in which any phase of the Pfaffian resulting from integrating the fermions
is neglected. We show numerical results indicating that this
approximation is likely adequate in the infinite volume continuum limit.
We have also conducted simulations employing temporal antiperiodic
boundary conditions for the fermions which further suppress the
phase fluctuations. 

We also show some new
results on the distribution of scalar fields in this
theory. We argue that the form of this distribution
is consistent with a lifting of the classical
moduli space via quantum fluctuations.
The final section addresses our conclusions.

\section{Twisted ${\cal N}=2$ SYM in two dimensions}
As discussed in \cite{n=2} the action of the 
the two dimensional ${\cal N}=2$ super Yang-Mills theory can be
written in the twisted or $Q$-exact form
\beq
S=\beta Q{\rm Tr}\int
d^2x\left(
\frac{1}{4}\eta[\phi,\phib]+2\chi_{12}F_{12}+\chi_{12}B_{12}+\psi_\mu D_\mu \phib\right)
\label{gfermion}
\eeq
where $Q$ is a scalar supercharge obtained by {\it twisting} the original
Majorana supercharges of the theory. The twist consists of decomposing all
fields under the action of a twisted rotation group. The latter
is obtained as the diagonal
subgroup of the direct product of the original (Euclidean) 
Lorentz symmetry with
an $SO(2)$ subgroup of the theory's R-symmetry.
In practice this means that we should treat the Lorentz and flavor
indices carried by all fields (and supercharges) 
as equivalent resulting in a $2\times 2$
matrix
structure for the fields. The twisted fields of the theory
correspond to the expansion coefficients when
this matrix is decomposed on
a basis of products of gamma matrices. 
Explicitly for the supercharge matrix $q$ this decomposition reads
\beq
q=QI+Q_\mu\gamma_\mu+Q_{12}\gamma_1\gamma_2\label{basis}\eeq
while a similar expression for the fermions allows the fermionic
content of the theory to be re-expressed in terms of the set
antisymmetric tensor fields $(\eta/2,\psi_\mu,\chi_{12})$. 
The action of the scalar supercharge on the component fields is\footnote{All
fields are to be thought of as {\it antihermitian} matrices in the
gauge group.}
\begin{eqnarray}
Q A_\mu&=&\psi_\mu\nonumber\\
Q \psi_\mu&=& -D_\mu\phi\nonumber\\
Q \chi_{12}&=&B_{12}\nonumber\\
Q B_{12}&=& [\phi,\chi_{12}]\nonumber\\
Q \phib&=&2\frac{\eta}{2}\nonumber\\
Q \frac{\eta}{2}&=&\frac{1}{2}[\phi,\phib]\nonumber\\
Q \phi&=&0
\end{eqnarray}
Notice that supersymmetry requires the introduction of $Q$-superpartners with
the same tensor structure as the fermions.
Carrying out the $Q$-variation on eqn.~\ref{gfermion}
and subsequently integrating over the field $B_{12}$ leads to the
action
\begin{eqnarray}
S&=&\beta {\rm Tr}\int d^2x\left(
\frac{1}{4}[\phi,\phib]^2-\frac{1}{4}\eta [\phi,\eta]-F_{12}^2-D_\mu \phi D_\mu \phib \right. \nonumber\\
&-&\left.\chi_{12} [\phi,\chi_{12}]-
2\chi_{12}\left(D_1\psi_2-D_2\psi_1\right)-\psi_\mu D_\mu\eta+\psi_\mu [\phib,\psi_\mu]\right)
\label{twist_sym_action}
\end{eqnarray}
The bosonic sector of this action is precisely the usual Yang-Mills
action while the fermionic sector constitutes a {\it K\"{a}hler-Dirac}
representation of the usual spinorial action
\cite{banks}. This can seen explicitly
by constructing a Dirac spinor out these twisted fields in the
following way
\beq
\lambda=\left(\begin{array}{c}
\frac{1}{2}\eta-i\chi_{12}\\
\psi_1-i\psi_2
\end{array}\right)\eeq
It is straightforward to see that the kinetic terms in 
\ref{twist_sym_action} can then be rewritten
in the Dirac form
\beq
\lambda^\dagger \gamma.D \lambda\eeq
where the gamma matrices are taken in the Euclidean chiral representation
\beq
\begin{array}{cc}
\gamma_1=\left(\begin{array}{cc}
0&1\\
1&0\end{array}\right)
&
\gamma_2=\left(\begin{array}{cc}
0&i\\
-i&0\end{array}\right) 
\end{array}
\eeq
In the same way the Yukawa interactions with the scalar fields can be written
\beq
\lambda^\dagger\frac{\left(1+\gamma_5\right)}{2}[\phib,\lambda]-
\lambda^\dagger\frac{\left(1-\gamma_5\right)}{2}[\phi,\lambda]\eeq
where $\gamma_5$ in this representation is
\beq
\gamma_5=\left(\begin{array}{cc}
1&0\\
0&-1\end{array}\right)\eeq
Thus the on-shell twisted action 
is nothing more than the usual ${\cal N}=2$ SYM action in two
dimensions. Indeed, in flat space the twisting process can be
thought of as simply a change of variables and is hence fully
equivalent to the usual formulation.\footnote{Notice that to make this correspondence and
obtain a bounded Euclidean action it is necessary to think of
$\phi$ as a complex matrix which is
(minus) the Hermitian conjugate of $\phib$.}

This
rewriting of the theory in terms of antisymmetric tensors has two
primary advantages -- it allows us to formulate the theory on
a curved space and as we will show in the next section
gives a natural starting point for
discretization. 

Finally it is worth pointing out that
the twisted theory possesses an additional $U(1)$ symmetry inherited from
the remaining R-symmetry of the model which is given by
\begin{eqnarray}
\psi_\mu\to e^{i\alpha}\psi_\mu\nonumber\\
\chi_{12}\to e^{-i\alpha}\chi_{12}\nonumber\\
\eta\to e^{-i\alpha}\eta\nonumber\\
\phi\to e^{2i\alpha}\phi\nonumber\\
\phib\to e^{-2i\alpha}\phib
\label{u1}
\end{eqnarray}
This symmetry can also be preserved under discretization and
ensures the absence of additive mass renormalizations in the theory.

\section{Lattice Action and Symmetries}
This theory may be discretized by mapping the
continuum rank $p$ antisymmetric
tensor fields to lattice fields living on $p$-cubes 
in a hypercubic lattice and replacing derivatives
with appropriate difference operators. 

The choice of difference operator
is very important to avoid fermion doubling -- one must
replace $\partial_\mu\to D^+_\mu$ if the derivative occurs in a curl-like
operation and $\partial\to D^-_\mu$ if the derivative belongs to
a divergence \cite{rabin}. 
Here, $D^+_\mu$ and $D^-_\mu$ refer to the usual forward and
backward difference operators respectively. Here gauge covariant
versions of
these difference operators must be used. We have employed the following
definitions \cite{aratyn}:
\begin{eqnarray}
D^+_\mu f(x) &=& U_\mu(x)f(x+\mu)-f(x)U_\mu(x)\nonumber\\
D^+_\mu f_\nu(x)&=& U_\mu(x)f_\nu(x+\mu)-f_\nu(x)U_\mu(x+\nu)
\end{eqnarray}
Notice that these definitions imply that the gauge transformations
of lattice scalar, vector and (antisymmetric) tensor fields are
\begin{eqnarray}
f(x)&\to& G(x)f(x)G^\dagger(x)\nonumber\\
f_\mu(x)&\to& G(x)f_\mu(x)G^\dagger(x+\mu)\nonumber\\
f_{\mu\nu}(x)&\to& G(x)f_{\mu\nu}(x)G^\dagger(x+\mu+\nu)
\end{eqnarray}
where $G(x)=e^{\phi(x)}$ is a lattice gauge transformation and all
fields are taken in the adjoint representation. 
From a lattice perspective these
transformations are very natural and correspond to thinking of each
field as living on a link running from the origin out to a vertex
on the unit hypercube.
The definitions of the backward difference operator follow by taking the
adjoint
\begin{eqnarray}
D^-_\mu f_{\mu\nu}(x)&=&f_{\mu\nu}(x)U^\dagger_\mu(x+\nu)-
U^\dagger_\mu f_{\mu\nu}(x-\mu)\nonumber\\
D^-_\mu f_\mu(x)&=&f_{\mu}(x)U^\dagger_\mu(x)-U^\dagger_\mu(x-\mu)f_\mu(x-\mu)
\end{eqnarray}
Notice that this lattice divergence has the merit of demoting the rank of
the lattice field as in the continuum and is consistent with the
lattice gauge transformation rules listed above.

The final step to constructing the lattice theory is to promote
each (anti)hermitian continuum field to a complex lattice field. This
allows $f$ and $f^\dagger$ to transform differently under gauge
transformations. This in turn is required if we are to
construct gauge invariant objects on the lattice. This doubling is actually
somewhat natural in a lattice theory with $p$-form lattice fields 
since
the underlying $p$-cube with $p>0$ has two orientations.
The complexification of
the vector potential $A_\mu^a(x)$ has the additional benefit of allowing
the fields $U(x)$ and $U^\dagger(x)$ to vary independently
under the twisted supersymmetry. 
In the end we will require 
the final path integral be taken along a contour where
$U^\dagger U=I$ and the imaginary parts of
the gauge field and the fermion fields vanish. This reality
condition will allow contact to be made with the usual
twisted continuum theory \cite{n=2,susysim}.

The $Q$-exact lattice action now takes the form
\begin{eqnarray}
S_L&=&\beta Q{\rm Tr}\sum_{x}\left(\frac{1}{4}\eta^\dagger(x)[\phi(x),\phib(x)]+
\chi^\dagger_{12}(x)\cF_{12}(x)+\chi_{12}(x)\cF_{12}(x)^\dagger\right.\nonumber\\
&+&\left.\frac{1}{2}\chi^\dagger_{12}(x)B_{12}(x)+\frac{1}{2}\chi_{12}(x)B^\dagger_{12}(x)+
\frac{1}{2}\psi^\dagger_\mu(x)D^+_\mu\phib(x)+\frac{1}{2}\psi_\mu(x)(D^+_\mu\phib(x))^\dagger\right)
\end{eqnarray}
This expression will also be $Q$-invariant if we can generalize the
continuum twisted supersymmetry transformations in such a way
that we preserve the property $Q^2=\delta_G^\phi$. The following transformations
do the job
\begin{eqnarray}
QU_\mu&=&\psi_\mu\nonumber\\
Q\psi_\mu&=&-D^+_\mu\phi\nonumber\\
Q\phi&=&0\nonumber\\
Q\chi_{12}&=&B_{12}\nonumber\\
QB_{12}&=&[\phi,\chi_{12}]^{(12)}\nonumber\\
Q\phib&=&\eta\nonumber\\
Q\eta&=&[\phi,\phib]
\end{eqnarray}
where the superscript notation indicates a {\it shifted} commutator
\beq
[\phi,\chi_{\mu\nu}]^{(\mu\nu)}=\phi(x)\chi_{\mu\nu}(x)-
\chi_{\mu\nu}(x)\phi(x+\mu+\nu)\eeq
These arise naturally when we consider the infinitesimal 
form of the gauge transformation
property of the plaquette field. Notice that gauge invariance also
dictates that we must use the covariant forward difference operator 
$D^+_\mu$ on the
right-hand side
of the $U_\mu$ variation. The lattice field strength
is given by $\cF_{\mu\nu}(x)=D^+_\mu U_\nu(x)$.

Carrying out this lattice $Q$-variation leads to the following expression
for the lattice action
\begin{eqnarray}
S_L&=&\beta{\rm Tr}\sum_x \left(
\frac{1}{4}[\phi(x),\phib(x)]^2-\frac{1}{4}\eta^\dagger(x)[\phi(x),\eta(x)]-
\chi^\dagger_{12}(x) [\phi(x),\chi_{12}(x)]^{(12)}+B^\dagger_{12}(x) B_{12}(x)\right.\nonumber\\
&+&B^\dagger_{12}(x)\cF_{12}(x)+B_{12}(x)\cF_{12}(x)^\dagger+
\frac{1}{2}(D^+_\mu \phi(x))^\dagger D^+_\mu \phib(x)+
\frac{1}{2}D^+_\mu \phi(x)(D^+_\mu\phib(x))^\dagger\nonumber\\
&-&\chi^\dagger_{12}(x)D^+_1\psi_2(x)+\chi^\dagger_{12}(x)D^+_2\psi_1(x))-
   \psi^\dagger_{2}(x)D^-_1\chi_{12}(x)+\psi^\dagger_1(x)D^-_2\chi_{12}(x)\nonumber\\
&-&\left.\frac{1}{2}\psi^\dagger_\mu(x)D^+_\mu\eta(x)-
         \frac{1}{2}\eta^\dagger(x) D^-_\mu \psi_\mu(x)+
\psi^\dagger_\mu(x) [\phib(x),\psi_\mu(x)]^{(\mu)}\right)
\label{twist_sym_latt_action}
\end{eqnarray}
Finally we must integrate out the multiplier fields $B_{12}$ and $B^\dagger_{12}$
resulting in the term
\beq
\beta {\rm Tr}\sum_x \cF_{12}(x)^\dagger\cF_{12}(x)\eeq
This can be written
\beq
\beta{\rm Tr}\sum_x \left(2I-U_P-U^\dagger_P\right)+
\beta{\rm Tr}\sum_x \left(M_{12}+M_{21}-2I\right)
\eeq
where \beq
U_P={\rm Tr}\left(U_1(x)U_2(x+1)U^\dagger_1(x+2)U^\dagger_2(x)\right)\eeq
resembles the
usual Wilson plaquette operator
and 
\beq M_{12}(x)=U_1(x)U^\dagger_1(x)U_2(x+1)U^\dagger_2(x+1)\eeq
Notice that the second term vanishes when the gauge
field is restricted to be unitary which is equivalent to 
requiring
${\rm Im}A_\mu(x)=0$. In this case the action is nothing more than
the usual Wilson gauge
action. 

Having constructed this complexified lattice theory possessing the
exact scalar supersymmetry we will subsequently truncate it to the
real line by setting the imaginary parts of all fields bar the
scalars to zero (the scalars are required to be (anti)hermitian conjugates
of each other as in the continuum). In \cite{susysim} we argued that
such a truncation should be valid for the Ward identities associated
with the scalar supersymmetry at least for sufficiently
large $\beta$ as a consequence of the $Q$-exact nature
of the action.
The numerical results presented in \cite{susysim} and in this
paper are consistent with this.

\section{Twisted Supersymmetries}
It is straightforward to construct the twisted supersymmetry transformations of
the component fields. It follows from
the matrix nature of the twist that the continuum fermion
kinetic term can be written in the form
\beq
S_F=\int d^2x{\rm Tr} \Psi^\dagger \gamma .D\Psi\eeq
where $\Psi$
corresponds to the
matrix form of the K\"{a}hler-Dirac field
\beq	
\Psi=\frac{\eta}{2}I+\psi_\mu\gamma_\mu+\chi_{12}\gamma_1\gamma_2\eeq
This term is clearly invariant under $\Psi\to \Psi\Gamma^i, i=1\ldots 4$ where
$\Gamma^i$ are the basis of products of gamma matrices introduced earlier
in eqn.~\ref{basis}.

Consider first the case $\Gamma^4=\gamma_1\gamma_2$.
In terms of the component fields the transformation $\Psi\to\Psi\Gamma^4$
effects a duality map
\begin{eqnarray}
\frac{\eta}{2}&\to& -\chi_{12}\nonumber\\
\chi_{12}&\to& \frac{\eta}{2}\nonumber\\
\psi_\mu&\to& -\epsilon_{\mu\nu}\psi_\nu
\end{eqnarray}
Such an operation clearly leaves the Yukawa terms invariant and trivially
all bosonic terms. It is thus a symmetry of the continuum action.
By combining such a transformation with the original 
action of the scalar
supercharge one derives a additional supersymmetry of the theory -- that
corresponding to the twisted supercharge $Q_{12}$.
Explicitly this supersymmetry will transform the component fields of
the continuum theory in
the following way
\begin{eqnarray}
Q_{12}A_\mu &=& -\epsilon_{\mu\nu}\psi_\nu\nonumber\\
Q_{12}\psi_\mu&=& -\epsilon_{\mu\nu} D_\nu \phi\nonumber\\
Q_{12}\chi_{12}&=&-\frac{1}{2}[\phi,\phib]\nonumber\\
Q_{12}B_{12}&=&[\phi,\frac{\eta}{2}]\nonumber\\
Q_{12}\phib&=&-2\chi_{12}\nonumber\\
Q_{12}\frac{\eta}{2}&=&B_{12}\nonumber\\
Q_{12}\phi&=&0
\end{eqnarray}
From the $Q$ and $Q_{12}$ transformations
it is straightforward to verify the following algebra holds 
\begin{eqnarray}
\{Q,Q\}&=&\{Q_{12},Q_{12}\}=\delta_\phi\nonumber\\
\{Q,Q_{12}\}&=&0
\end{eqnarray}
where $\delta_\phi$ denotes an infinitessimal gauge transformation with
parameter $\phi$. This allows us to construct strictly nilpotent
symmetries $\hat{Q}_{\pm}=Q\pm iQ_{12}$ in the continuum theory
corresponding to using the (anti)self-dual components of the
original \KD field.

In the same way we can try to build an additional supersymmetry
by combining the invariance of the fermion kinetic term 
under $\Psi\to\Psi\Gamma^1$ with the existing scalar supersymmetry.
This effects the following transformation of fermion fields:
\begin{eqnarray}
\frac{\eta}{2}&\to&\psi_1\nonumber\\
\chi_{12}&\to& -\psi_2\nonumber\\
\psi_1&\to& \frac{\eta}{2}\nonumber\\
\psi_2&\to&-\chi_{12}
\end{eqnarray}
However, the Yukawas and bosonic terms are only invariant under
such a transformation if we simultaneously make the
transformation $\phi\to-\phib$.
The resultant explicit action of $Q_1$ on the component fields is given
by
\begin{eqnarray}
Q_{1}A_1 &=& \frac{\eta}{2}\nonumber\\
Q_{1}A_2&=&-\chi_{12}\nonumber\\
Q_{1}\psi_1&=&-\frac{1}{2}[\phi,\phib]\nonumber\\
Q_{1}\psi_2&=&-B_{12}\nonumber\\
Q_{1}\chi_{12}&=&-D_2\phib\nonumber\\
Q_{1} B_{12}&=&[\phib,\psi_2]\nonumber\\
Q_{1}\phib&=&0\nonumber\\
Q_{1}\frac{\eta}{2}&=& D_1\phib\nonumber\\
Q_{1}\phi&=&-2\psi_1
\end{eqnarray}
Similarly the supersymmetry associated with $\Gamma^2$ is given by
\begin{eqnarray}
Q_{2} A_1&=&\chi_{12}\nonumber\\
Q_{2} A_2&=&\frac{\eta}{2}\nonumber\\
Q_{2} \psi_1&=&B_{12}\nonumber\\
Q_{2} \psi_2&=&-\frac{1}{2}[\phi,\phib]\nonumber\\
Q_{2} \chi_{12}&=&D_1\phib\nonumber\\
Q_{2} B_{12}&=&-[\phib,\psi_1]\nonumber\\
Q_{2} \phib&=&0\nonumber\\
Q_{2} \frac{\eta}{2}&=&D_2\phib\nonumber\\
Q_{2} \phi&=&-2\psi_2
\end{eqnarray}
Again, we can verify the following algebra holds
\begin{eqnarray}
\{Q_1,Q_1\}&=&\{Q_2,Q_2\}=\delta_{-\phib}\nonumber\\
\{Q_1,Q_2\}&=&0
\end{eqnarray}
with $\delta_{-\phib}$ a corresponding gauge transformation with parameter
$-\phib$. This allows us to construct yet another pair of nilpotent
supercharges in the continuum theory $\overline{Q}_\pm=Q_1\pm iQ_2$. 

It is interesting to check also the anticommutators of these new charges
$\hat{Q}_\pm$ and $\overline{Q}_\pm$. It is a straightforward exercise to
verify the following algebra holds {\it on-shell} 
\begin{eqnarray}
\{\hat{Q}_+,\overline{Q}_-\}&=&\{\hat{Q}_-,\overline{Q}_+\}=0\nonumber\\
\{\hat{Q}_+,\overline{Q}_+\}&=&4(D_1+iD_2)\nonumber\\
\{\hat{Q}_-,\overline{Q}_-\}&=&4(D_1-iD_2)
\label{anti}
\end{eqnarray}
As an example consider $\{\hat{Q}_+,\overline{Q}_+\}\psi_1$
\beq
\{\hat{Q}_+,\overline{Q}_+\}=\{Q,Q_1\}-\{Q_{12},Q_2\}+i(\{Q_{12},Q_1\}+\{Q,Q_2\}
\eeq
Using the component transformations listed above
the relevant anticommutators are
\begin{eqnarray}
\{Q,Q_1\}\psi_1&=&2D_1\psi_1\nonumber\\
\{Q,Q_2\}\psi_1&=&2D_1\psi_2+2[\phi,\chi_{12}]\nonumber\\
\{Q_{12},Q_1\}\psi_1&=&2D_2\psi_1\nonumber\\
\{Q_{12},Q_2\}\psi_1&=&2D_2\psi_2+[\phi,\eta]
\end{eqnarray}
Thus we find
\beq
\{\hat{Q}_+,\overline{Q}_+\}=2D_1\psi_1-2D_2\psi_2-[\phi,\eta]+
                           i(2D_1\psi_2+2D_2\psi_1+2[\phi,\chi_{12}])\eeq
Using the equations of motion
\begin{eqnarray}
-2D_1\psi_1-2D_2\psi_2-[\phi,\eta]&=&0\nonumber\\
-2D_2\psi_1+2D_1\psi_2+2[\phi,\chi_{12}]&=&0
\end{eqnarray}
we can easily verify the second line of eqn.~\ref{anti}. Notice that
from these new charges $\hat{Q}_{\pm}$, $\overline{Q}_\pm$ 
we can build spinorial supercharges of the form
\beq\left(\begin{array}{c}
\hat{Q}_+\\
\overline{Q}_-
\end{array}\right)
\eeq
in which case the algebra given in eqn.~\ref{anti} represents
the usual supersymmetry algebra in a chiral basis (up to a gauge
transformation).

Finally we will be interested in Ward identities which can be derived for
a general operator $O$ and take the form
\beq
<Q^iO>=0\qquad i=1\ldots 4\eeq
for each of the four supersymmetries. To get nontrivial results the operator
$O$ must be gauge invariant and have $U(1)$ charge $-1$.

\section{Numerical Results}
We have implemented the RHMC dynamical fermion algorithm to simulate this
lattice theory -- for details we refer the reader to \cite{susysim}.
We have examined lattices with $L^2$ geometry with $L$ in the range
$3-8$ and a variety of lattice couplings. Simulations
have been done employing both periodic boundary
conditions and thermal boundary conditions corresponding to
enforcing antiperiodicity on the fermions in the temporal
direction. 
Following \cite{susysim} we have worked in the phase quenched
ensemble -- the Pfaffian arising from integration over the
twisted fermions is replaced by its absolute value within the
Monte Carlo simulation. We do, however, monitor the
phase and in the next section will discuss the 
the results of reweighting our measured observables
with those phase fluctuations.
Typically, our results
derive from $O(10^{3-4})$ trajectories for both $SU(2)$ and $SU(3)$ theories
where a single trajectory corresponds to $\tau=1$ units of classical
dynamics time.
\subsection{Scalar supersymmetry}
Consider first a series of Ward identities associated with the
exact lattice supersymmetry $Q$. The simplest of these corresponds
to the statement $<S>=-\frac{\partial \ln{Z}}{\partial\beta}=0$ 
reflecting the $Q$-exact nature of the
twisted action.
If we integrate out the twisted fermions and the auxiliary
field $B_{12}$ we find the following expression for the 
partition function
\beq
Z=\beta^{4N_GN_s/2}\beta^{-N_GN_s/2}
\int D\phi D\phib DU e^{-\beta S_B(\phi,\phib,U)}{\rm det}(M(\Phi))\eeq
where $N_s$ is the number of sites and $N_G$ the number of generators
of the gauge group. The first prefactor arises from the fermion
integration while the second derives from the gaussian
integration over the auxiliary field.
From this we find the following condition on the mean bosonic
action as a consequence of the scalar supersymmetry
\beq
\frac{2}{3N_GN_s}<\beta S_B>=1\eeq
To examine this quantity in the continuum limit we must know how to
scale the lattice coupling $\beta$ with the number of lattice
points. Clearly the physics of the model is determined by the
{\it dimensionless} coupling $\mu=g^2A$ where $A$ is the physical
area. If we simply equate this to the corresponding lattice
quantity we find the scaling
\[\beta=\frac{L^2}{\mu}\]
The continuum limit is thus gotten from this equation taking $L\to\infty$
while holding $\mu$ fixed. Notice that the infinite volume limit 
corresponds then to taking the subsequent limit $\mu\to\infty$.

Figure~\ref{action} shows a plot of this quantity for the $SU(3)$ gauge
group for $\mu=0.25$ and $\mu=5.0$ as the continuum limit
is approached with $L=3,4,5$.
While small deviations of order $1.5$\% are seen from the
theoretical result based on exact supersymmetry on the smallest lattice with
$\mu=0.25$ these appear to diminish as the continuum limit is taken.
Furthermore, at $\mu=5.0$ the data appears consistent with exact supersymmetry
for all lattice sizes (the maximum deviation being $0.5$\%). 
\begin{table}\begin{center}
\begin{tabular}{||l|l|l|l|l||}
\hline
$$ & \multicolumn{2}{c|}{$SU(2)$}&\multicolumn{2}{c|}{$SU(3)$}         \\\hline
$L$   & $B$     & $F$          & $B$         & $F$                 \\\hline
$3$ & $0.031(2)$  & $-0.032(2)$   & $0.13(2)$  & $-0.13(2)$          \\\hline
$4$  & $0.0087(8)$  & $-0.0094(6)$   & $0.053(9)$  & $-0.055(9)$          \\\hline
$5$  & $0.0032(5)$  & $-0.0038(6)$   & $0.0091(9)$    & $-0.0109(8)$ \\\hline
\end{tabular}\label{tab1}
\caption{$QO_1$ vs $L$ for $SU(2)$, $SU(3)$ and $\mu=0.25$}
\end{center}\end{table}
\begin{table}\begin{center}
\begin{tabular}{||l|l|l|l|l||}
\hline
$$ & \multicolumn{2}{c|}{$SU(2)$}&\multicolumn{2}{c|}{$SU(3)$}         \\\hline
$L$   & $B$     & $F$          & $B$         & $F$                 \\\hline
$3$ & $0.0334(7)$  & $-0.0363(5)$   & $0.077(3)$  & $-0.083(4)$          \\\hline
$4$  & $0.0215(3)$  & $-0.0222(2)$   & $0.049(3)$  & $-0.051(2)$          \\\hline
$5$  & $0.016(1)$  & $-0.0146(1)$   & $0.036(1)$    & $-0.0382(2)$           \\\hline
\end{tabular}\label{tab2}
\caption{$QO_2$ vs $L$ for $SU(2)$, $SU(3)$ and $\mu=0.25$}
\end{center}\end{table}
We can easily derive other Ward identities associated with
the $Q$-symmetry by taking the $Q$-variation of the following
operators $O_1(x)=\eta^\dagger(x)[\phi(x),\eta(x)]$, 
$O_2(x)=\chi^\dagger_{12}(x)F_{12}(x)$ and
$O_3(x)=\psi^\dagger_\mu(x) D_\mu\phib(x)$. In practice we have examined
the integrated quantities $\sum_x O_i(x)$ and estimated the latter
by selecting a source point $x$ randomly on each Monte Carlo configuration.
Such a procedure minimizes the statistical errors for a fixed amount
of computation.
After $Q$-variation we find
\begin{eqnarray}
QO_1&=&[\phi,\phib]^2-\eta^\dagger[\phi,\eta]\nonumber\\
QO_2&=&F^\dagger_{\mu\nu}(x)F_{\mu\nu}(x)-
\chi^\dagger_{\mu\nu}D^+_{\left[\mu\right.}\psi_{\left.\nu\right]}\nonumber\\
QO_3&=&-D^+_\mu\phi D^+_\mu\phib-\psi^\dagger_\mu D^+_\mu\eta-\psi^\dagger_\mu[\psib_\mu,\phi]
\end{eqnarray}
The results for the expectation values of these $Q$-variations 
as a function of lattice size $L=3,4,5$ are shown in tables
1,2 and 3. for $\mu=0.25$
and gauge groups $SU(2)$ and
$SU(3)$. Similar data at $\mu=10.0$ (for $SU(2)$) and $\mu=5.0$
(for $SU(3)$) are shown in tables 4,5, and 6.
We denote the bosonic contribution to the Ward identity by $B$ and
the real part of the fermionic contribution by $F$ -- the imaginary part is
always small and statistically consistent with zero. 
\begin{table}\begin{center}
\begin{tabular}{||l|l|l|l|l||}
\hline
$$ & \multicolumn{2}{c|}{$SU(2)$}&\multicolumn{2}{c|}{$SU(3)$}         \\\hline
$L$   & $B$     & $F$          & $B$         & $F$                 \\\hline
$3$ & $0.085(3)$  & $-0.0808(3)$   & $0.21(1)$  & $-0.2172(5)$          \\\hline
$4$  & $0.046(2)$  & $-0.0460(3)$   & $0.11(1)$  & $-0.1232(4)$       \\\hline
$5$  & $0.029(3)$  & $-0.0295(1)$   & $0.081(6)$    & $-0.0790(4)$\\\hline
\end{tabular}\label{tab3}
\caption{$QO_3$ vs $L$ for $SU(2)$, $SU(3)$ and $\mu=0.25$}
\end{center}\end{table}
\begin{table}\begin{center}
\begin{tabular}{||l|l|l|l|l||}
\hline
$$ & \multicolumn{2}{c|}{$SU(2)$}&\multicolumn{2}{c|}{$SU(3)$}         \\\hline
$L$   & $B$     & $F$          & $B$         & $F$                 \\\hline
$3$ & $2.7(2)$  & $-2.8(3)$   & $3.8(1)$  & $-3.9(1)$          \\\hline
$4$  & $0.81(3)$  & $-0.91(3)$   & $1.12(4)$  & $-1.16(3)$          \\\hline
$5$  & $0.39(1)$  & $-0.44(1)$   & $0.50(2)$    & $-0.55(2)$   \\\hline
\end{tabular}\label{tab4}
\caption{$QO_1$ vs $L$ for $SU(2)$ ($\mu=10.0$) and $SU(3)$ ($\mu=5.0$)}
\end{center}\end{table}
To give a more graphical illustration of the
presence of the exact scalar supersymmetry $Q$ we show
in figure~\ref{o1} plots of $B+F$ versus $L$ for $\mu=0.25,5.0$
and gauge group $SU(3)$.
\begin{table}\begin{center}
\begin{tabular}{||l|l|l|l|l||}
\hline
$$ & \multicolumn{2}{c|}{$SU(2)$}&\multicolumn{2}{c|}{$SU(3)$}         \\\hline
$L$   & $B$     & $F$          & $B$         & $F$                 \\\hline
$3$ & $1.05(8)$  & $-1.07(8)$   & $1.27(4)$  & $-1.31(4)$          \\\hline
$4$  & $0.755(9)$  & $-0.777(6)$   & $1.00(2)$  & $-1.003(7)$          \\\hline
$5$  & $0.51(3)$  & $-0.53(2)$   & $0.67(1)$    & $-0.686(5)$           \\\hline
\end{tabular}\label{tab5}
\caption{$QO_2$ vs $L$ for $SU(2)$ ($\mu=10.0$) and $SU(3)$ ($\mu=5.0$)}
\end{center}\end{table}
\begin{table}\begin{center}
\begin{tabular}{||l|l|l|l|l||}
\hline
$$ & \multicolumn{2}{c|}{$SU(2)$}&\multicolumn{2}{c|}{$SU(3)$}         \\\hline
$L$   & $B$     & $F$          & $B$         & $F$                 \\\hline
$3$ & $3.40(8)$  & $-3.24(2)$   & $4.51(5)$  & $-4.36(1)$          \\\hline
$4$  & $1.87(2)$  & $-1.812(3)$   & $2.46(4)$  & $-2.454(2)$          \\\hline
$5$  & $1.165(9)$  & $-1.162(2)$   & $1.570(30)$    & $-1.590(10)$           \\\hline
\end{tabular}\label{tab6}
\caption{$QO_3$ vs $L$ for $SU(2)$ ($\mu=10.0$) and $SU(3)$ ($\mu=5.0$)}
\end{center}\end{table}
In general we see that these Ward identities are rather well satisfied
for all lattice sizes and certainly as we move
toward the continuum limit. 

\subsection{The broken supersymmetries}
In this section we examine Ward identities corresponding to the
supersymmetries which are broken after discretization. Referring to the
previous section we see that these supercharges in general will transform
a field living on one link to a neighboring link in a manner 
similar to the link constructions used in \cite{noboru}.
This in turn implies that the variation of any closed, gauge invariant
loop will vary into an open gauge variant loop. The latter will
{\it automatically} have vanishing expectation value due to gauge
invariance. Thus the set of operators whose variation under
one of these link supersymmetries yields
a non-trivial Ward identity is rather small and is further constrained
by the need to obtain singlets under the additional $U(1)$ symmetry
described in eqn.~\ref{u1}.
We have examined the following
ones
\begin{eqnarray}
Q_{12}O_4&=&Q_{12}\left(\chi^\dagger_{12}[\phi,\phib]\right)=
\frac{1}{4}[\phi,\phib]^2-\chi^\dagger_{12}[\phi,\chi_{12}]\\
Q_1 O_5&=&Q_1\left(\psi^\dagger_1[\phi,\phib]\right)=\frac{1}{4}[\phi,\phib]^2-
\psi^\dagger_1[\psi_1,\phib]\\
Q_2 O_6&=&Q_2\left(\psi_2[\phi,\phib]\right)=\frac{1}{4}[\phi,\phib]^2-
\psi^\dagger_2[\psi_2,\phib]
\end{eqnarray}
They correspond to terms in the lattice action. Notice that they imply that
the expectation values of
different Yukawa interactions should be equal. Tables 7 and 8
show data for $SU(2)$ and $SU(3)$ and lattice sizes $L=3,4,5$
at $\mu=0.25$. Similar data for $SU(2)$ at $\mu=10.0$
and $SU(3)$ at $\mu=5.0$ are shown in tables 9 and 10.
The results for the Ward identity $Q_2O_6$ are very similar
to those for $Q_1$ and we omit them. Some of this
data is also shown graphically in figures \ref{o4} and \ref{o5} 
which again plot $B+F$ against $L$ for 
$SU(3)$ at $\mu=0.25,5.0$ for the operators $O_4$ and $O_5$.
\begin{table}\begin{center}
\begin{tabular}{||l|l|l|l|l||}
\hline
$$ & \multicolumn{2}{c|}{$SU(2)$}&\multicolumn{2}{c|}{$SU(3)$}         \\\hline
$L$   & $B$     & $F$          & $B$         & $F$                 \\\hline
$3$ & $0.077(6)$  & $-0.0053(5)$   & $0.032(4)$  & $-0.028(4)$          \\\hline
$4$  & $0.0022(2)$  & $-0.0013(1)$   & $0.013(3)$  & $-0.012(3)$          \\\hline
$5$  & $0.0008(1)$  & $-0.0004(1)$   & $0.0023(2)$    & $-0.0018(2)$           \\\hline
\end{tabular}\label{tab7}
\caption{$Q_{12}O_4$ vs $L$ for $SU(2)$, $SU(3)$
for $\mu=0.25$}
\end{center}\end{table}
\begin{table}\begin{center}
\begin{tabular}{||l|l|l|l|l||}
\hline
$$ & \multicolumn{2}{c|}{$SU(2)$}&\multicolumn{2}{c|}{$SU(3)$}         \\\hline
$L$   & $B$     & $F$          & $B$         & $F$                 \\\hline
$3$ & $0.077(6)$  & $-0.066(6)$   & $0.032(4)$  & $-0.030(4)$          \\\hline
$4$  & $0.0022(2)$  & $-0.018(1)$   & $0.013(3)$  & $-0.013(2)$          \\\hline
$5$  & $0.0008(1)$  & $-0.0007(1)$   & $0.0023(2)$    & $-0.0023(2)$           \\\hline
\end{tabular}\label{tab8}
\caption{$Q_1O_5$ vs $L$ for $SU(2)$, $SU(3)$ for $\mu=0.25$}
\end{center}\end{table}
\begin{table}\begin{center}
\begin{tabular}{||l|l|l|l|l||}
\hline
$$ & \multicolumn{2}{c|}{$SU(2)$}&\multicolumn{2}{c|}{$SU(3)$}         \\\hline
$L$   & $B$     & $F$          & $B$         & $F$                 \\\hline
$3$ & $0.67(8)$  & $-0.6(1)$   & $0.95(3)$  & $-0.92(4)$          \\\hline
$4$  & $0.204(7)$  & $-0.161(6)$   & $0.28(1)$  & $-0.25(1)$          \\\hline
$5$  & $0.099(3)$  & $-0.072(2)$   & $0.124(6)$    & $-0.114(5)$           \\\hline
\end{tabular}\label{tab9}
\caption{$Q_{12}O_4$ vs $L$ for $SU(2)$ ($\mu=10.0$) and $SU(3)$ ($\mu=5.0$)}
\end{center}\end{table}

\begin{table}\begin{center}
\begin{tabular}{||l|l|l|l|l||}
\hline
$$ & \multicolumn{2}{c|}{$SU(2)$}&\multicolumn{2}{c|}{$SU(3)$}         \\\hline
$L$   & $B$     & $F$          & $B$         & $F$                 \\\hline
$3$ & $0.67(8)$  & $-0.64(7)$   & $0.95(3)$  & $-0.95(3)$          \\\hline
$4$  & $0.204(7)$  & $-0.193(6)$   & $0.28(1)$  & $-0.26(9)$          \\\hline
$5$  & $0.099(3)$  & $-0.092(3)$   & $0.124(6)$    & $-0.123(5)$      \\\hline
\end{tabular}\label{tab10}
\caption{$Q_1O_5$ vs $L$ for $SU(2)$ ($\mu=10.0$) and $SU(3)$ ($\mu=5.0$)}
\end{center}\end{table}

We see no statistically significant evidence for breaking of these
Ward identities in these tables {\it except} for the case of $Q_{12}O_4$
and gauge group $SU(2)$. In this case the breaking appear worse
for small $\mu$ and appear to survive the continuum limit
$L\to\infty$.
We conjecture that the explanation for these breakings lies in the
use of the phase quenched approximation. In the next section we
examine this issue more carefully and find evidence that the
problem of phase fluctuations gets worse at small $\mu$. Moreover,
since the latter are driven by near zero modes of the fermion
operator it is also plausible that the effect is enhanced for
small $N$ since the number of fermionic zero modes scales as
$N$ while the number of non zero modes varies as $N^2$.

\subsection{Phase quenching}
In this section we try to quantify the magnitude of any corrections to
the Ward identities due to phase quenching.
Initially we focus on lattices employing periodic boundary conditions
on all fields. Tables 11 and 12
show the cosine of the phase $<\cos{\alpha}>$
for the $SU(2)$ theory
at $\mu=0.25$ and $\mu=10.0$ respectively.
\DOUBLETABLE
{\begin{tabular}{||l|l||}
\hline
$L$  & $<\cos{\alpha}>$ \\\hline   
$3$  & $0.14(1)$ \\\hline
$4$  & $0.13(2)$ \\\hline
$5$  & $0.14(4)$ \\\hline
\end{tabular}\label{tab11}}
{\begin{tabular}{||l|l||}
\hline
$L$ & $<\cos{\alpha}>$ \\\hline
$3$ & $0.33(6)$ \\\hline
$4$ & $0.37(1)$ \\\hline
$5$ & $0.27(1)$ \\\hline
\end{tabular}\label{tab12}}
{$SU(2)$ $\mu=0.25$}
{$SU(2)$ $\mu=10.0$}
Notice that discretization effects appear to be small and that the mean value
increases with increasing $\mu$. A plot of the distribution $P(\cos{\alpha})$
for $SU(2)$ from simulations at $L=3$ and $\mu=0.25$
is shown in figure~\ref{pbcmu0.25}. A similar picture for $\mu=10.0$
is plotted in figure~\ref{pbcmu10.0}. 
From this figure we see that the
distribution possesses two peaks - one at $\cos{\alpha}=1$ and another
at $\cos{\alpha}=-1$. Thus the measured Pfaffian is predominately
real but has an indefinite sign. From the observed weak dependence of
$<\cos{\alpha}>$ on $L$ and direct observation of the distribution
$P(\cos{\alpha})$ at increasing $L$ we conclude that these phase
fluctuations survive the continuum limit for any fixed $\mu$. 
Notice though that the 
height of the $\cos{\alpha}=-1$ peak {\it decreases} relative to
the $\cos{\alpha}=1$ peak as $\mu$
increases which is responsible 
for the observed increase of $<\cos{\alpha}>$ with $\mu$.  We conjecture
that the phase becomes concentrated
at $\cos{\alpha}=1$ in the infinite volume
limit corresponding to $\mu\to\infty$. In this limit then the
phase quenched approximation would be exact.

Furthermore, we conjecture that 
neglect of these phase fluctuation is the origin of
the breaking of the $Q_{12}$ Ward identity observed in the previous
section for the $SU(2)$ theory
at small $\mu$. We have attempted to check this by examining the
reweighted values of the bosonic action. The data is shown in table 13
which shows $<S_B>$ and its reweighted cousin $S_B^R$
as a function of $\mu$ from simulations of
the $SU(2)$ theory with $L=3$.
\TABLE{
\begin{tabular}{||l|l|l||}
\hline
$\mu$  & $S_B$       & $S^R_B$       \\\hline
$0.25$ & $0.977(1)$  & $0.989(70)$   \\\hline
$1.0$  & $0.981(1)$  & $0.995(45)$   \\\hline
$5.0$  & $0.992(2)$  & $1.000(32)$   \\\hline
$10.0$ & $0.999(2)$  & $1.003(160)$  \\\hline
\end{tabular}\label{tab13}
\caption{$S_B$ and reweighted $S^R_B$ vs $\mu$ for $SU(2)$ and $L=3$}}
While the measured value of the reweighted action
has very large errors there is an indication that it lies closer to the
exact value relative to the
unreweighted action at small $\mu$. Notice again though that
even the naive mean action appears to approach the expected
theoretical value for sufficiently large $\mu$.

To summarize we have observed non-trivial fluctuations in the phase of
the Pfaffian which appear to survive
the continuum limit for any finite continuum coupling $\mu$. 
These phase fluctuations appear to become less important in the
infinite volume limit corresponding to $\mu\to\infty$.

We conjecture
that this non-trivial phase is associated primarily with the fluctuations of
lowest lying eigenvalues of the fermion operator. We have directly
observed such states which correspond to the superpartners of the bosonic
zero modes associated with the classical moduli space. One simple way to
try to reduce the problem is to consider the theory
at finite temperature by employing {\it 
antiperiodic} temporal boundary conditions on the fermions. We consider this
in detail in the next section.
\subsection{Thermal boundary conditions}
In tables 14 and 15
we show the
measured values of $<\cos{\alpha}>$ versus $L$ for couplings
$\mu=0.25$ and $\mu=10.0$ from thermal
simulations using gauge group $SU(2)$. 
Notice that the use of thermal boundary conditions does indeed push the
distribution toward $\cos{\alpha}=1$. This is
consistent with the effect being driven primarily by the
near zero modes whose eigenvalues are lifted to $O(1/L)$ 
at finite temperature. The distribution for $\mu=10.0$ in the
$SU(2)$ theory at $L=3$ is shown in figure\ref{apbcmu10.0}. In this
case the distribution is essentially zero away from a very sharp
peak at $\cos{\alpha}=1$ whose width is only $O(0.01)$. Notice, however,
that this peak broadens as the continuum limit is approach and the
expectation value of $\cos{\alpha}$ subsequently falls.
\DOUBLETABLE
{\begin{tabular}{||l|l||}
\hline
$L$  & $<\cos{\alpha}>$ \\\hline   
$3$  & $0.86(8)$ \\\hline
$4$  & $0.86(3)$ \\\hline
$5$  & $0.72(5)$ \\\hline
\end{tabular}\label{tab14}}
{\begin{tabular}{||l|l||}
\hline
$L$ & $<\cos{\alpha}>$ \\\hline
$3$ & $0.95(2)$ \\\hline
$4$ & $0.79(5)$ \\\hline
$5$ & $0.50(1)$ \\\hline
\end{tabular}\label{tab15}}
{$SU(2)$ apbc $\mu=0.25$}
{$SU(2)$ apbc $\mu=10.0$}
We have repeated the calculations of the supersymmetric Ward
identities for the thermal case and the results for $QO_1$,
$Q_{12}O_4$ and $Q_1O_5$ are shown
in tables 16, 17 and 18
at couplings $\mu=10.0$ for $SU(2)$ and $\mu=5.0$ for
$SU(3)$. 
\begin{table}\begin{center}
\begin{tabular}{||l|l|l|l|l||}
\hline
$$ & \multicolumn{2}{c|}{$SU(2)$}&\multicolumn{2}{c|}{$SU(3)$}         \\\hline
$L$   & $B$     & $F$          & $B$         & $F$                 \\\hline
$3$ & $4.2(2)$  & $-4.2(2)$   & $5.2(3)$  & $-5.3(2)$          \\\hline
$4$  & $1.8(2)$  & $-1.8(2)$   & $2.1(1)$  & $-2.1(1)$          \\\hline
$5$  & $0.53(2)$  & $-0.55(2)$   & $0.90(8)$    & $-0.93(8)$ \\\hline
\end{tabular}\label{tab16}
\caption{$QO_1$ for $SU(2)$ with apbc ($\mu=10.0$) and $SU(3)$ ($\mu=5.0$)}
\end{center}\end{table}
\begin{table}\begin{center}
\begin{tabular}{||l|l|l|l|l||}
\hline
$$ & \multicolumn{2}{c|}{$SU(2)$}&\multicolumn{2}{c|}{$SU(3)$}         \\\hline
$L$   & $B$     & $F$          & $B$         & $F$                 \\\hline
$3$ & $1.05(5)$  & $-1.03(6)$   & $1.30(6)$  & $-1.29(6)$          \\\hline
$4$  & $0.45(5)$  & $-0.42(6)$   & $0.53(3)$  & $-0.50(4)$          \\\hline
$5$  & $0.132(6)$  & $-0.101(6)$   & $0.22(2)$    & $-0.21(2)$ \\\hline
\end{tabular}\label{tab17}
\caption{$Q_{12}O_4$ for $SU(2)$ with apbc ($\mu=10.0$) and $SU(3)$ ($\mu=5.0$)}
\end{center}\end{table}
\begin{table}[h]\begin{center}
\begin{tabular}{||l|l|l|l|l||}
\hline
$$ & \multicolumn{2}{c|}{$SU(2)$}&\multicolumn{2}{c|}{$SU(3)$}      \\\hline
$L$   & $B$     & $F$          & $B$         & $F$                 \\\hline
$3$ & $1.05(5)$  & $-1.03(6)$   & $1.30(6)$  & $-1.31(6)$          \\\hline
$4$  & $0.45(5)$  & $-0.44(5)$   & $0.53(3)$  & $-0.51(3)$         \\\hline
$5$  & $0.132(6)$  & $-0.119(5)$   & $0.22(2)$    & $-0.22(2)$     \\\hline
\end{tabular}\label{tab18}
\caption{$Q_1O_5$ for $SU(2)$ with apbc ($\mu=10.0$) and $SU(3)$ ($\mu=5.0$)}
\end{center}\end{table}
For the thermal runs we
are interested primarily in the data for large
coupling $\mu$ which is now related to the inverse temperature - thus
large $\mu$ corresponds to the theory at low temperature and/or large
spatial volume. 
The data is also illustrated graphically in figure~\ref{apbcward} for
the case of $SU(3)$. 
In this regime we find that the all the supersymmetric
Ward identities computed within the phase quenched approximation
are satisfied within statistical errors - thus we conclude that
the use of thermal boundary conditions may indeed be quite
useful in reducing phase fluctuation problems and eliminating the
need for reweighting even away from the infinite volume limit.

In the absence of spontaneous supersymmetry breaking expectation
values computed in
the thermal system should
approach those of the periodic/zero temperature theory for sufficiently
low temperatures. No additional fine tuning should be required and this
is seen in our data for the Ward identities.

\subsection{Rotational invariance}
To test for a restoration of rotational invariance as $\beta$ is increased
and the continuum limit is approached we have examined the two-point
function
\beq
G(x,y)=<B(0,0)B(x,y)>\eeq
where $B(x,y)=[\phi(x,y),\phib(x,y)]$ and $x$ and $y$ are integer lattice
coordinates. If the theory is rotationally invariant we would expect
this correlator to be a function of just
the radial distance $r=\sqrt{x^2+y^2}$. Figure~\ref{corrsu2}. shows a
plot of this function for the $SU(2)$ theory at $\beta=8.0$ on a
$8\times 8$ lattice.
The
points lie within errors on a single curve lending support to
the idea that at least the scalar
sector of the theory is rotationally invariant at large distance.
If all supersymmetries are restored for sufficiently small
lattice spacing we would then expect the fermionic sector to
also be rotationally invariant in the continuum limit.
Fig~\ref{corrsu3} shows similar data for $SU(3)$ at $\beta=4.0$ and
leads to similar conclusions.

\subsection{Fluctuations of the scalar fields}
The classical vacua of this theory allow for
any set of scalars which are constant over the lattice and
satisfy 
\beq[\phi,\phib]=0\eeq
In the case of $SU(2)$ and using the parameterization $\phi=\phi_1+i\phi_2$
we find vacuum states of the form 
\beq \phi_1=(A,0,0)\qquad \phi_2=(B,0,0)\eeq
together with
global $SU(2)$ rotations of this configuration.
Thus we have a classical vacuum state for any value of
$A$ and $B$. In the case of $SU(3)$ the general vacuum state can be
realized by taking
the fields $\phi_1$ and $\phi_2$ as constant diagonal matrices
with arbitrarily large matrix elements. Generalization to arbitrary
$SU(N)$ the space of vacuum solutions is referred to as the
classical moduli space.
The presence of such a non-trivial
moduli space corresponds to the existence of flat directions 
in the theory which have the potential to 
lead to divergences when inserted
into a path integral. Notice, however, that the fermion
operator develops zero modes precisely along these bosonic
flat directions and thus quantum
fluctuations of the fermions may suppress their contribution. In addition,
entropic effects at large $N$
may also play a role in removing potential divergences since
the number of classical zero modes scales like $N$ while the number of
non-zero
modes scales like $N^2$.
It then becomes a dynamical
question as to whether the theory retains a non-trivial moduli
space at the quantum level. 

To examine this issue we have
measured the probability distribution of the scalar eigenvalues seen in
the simulation. 
Figure~\ref{su2}. shows a plot of this distribution of scalar field
eigenvalues for $SU(2)$ obtained from high statistics runs ($10^5$ Monte
Carlo trajectories) on a 
$3\times 3$ lattice. We have mapped all data to positive values by
exploiting the $\lambda\to -\lambda$ symmetry of the
action.
The two sets of points correspond to $\beta=0.5$
and $\beta=4.0$. Contrary to the classical analysis it appears that
the most probable scalar field configurations lie close to the
origin in field space and this effect is enhanced as the coupling
$\beta$ is increased. 
A similar situation is seen for $SU(3)$ in figure~\ref{su3}. --
the main difference being that there are now
three peaks in this distribution (only two are 
seen after the mapping to the positive $\lambda$ axis)
with a new peak appearing at the origin. This pattern
repeats at larger $N$ -- in the case of 
$SU(N)$ we have observed
that this probability distribution possesses $N$ peaks \cite{tobyme}.
Since the scalars in this theory 
arise from dimensional reduction of gauge fields
in four dimensional ${\cal N}=1$ Yang-Mills it seems likely that
these $N$ peaks in the scalar distribution are analogous to
the $N$ peaks appearing
in the distribution of the expectation value of the Polyakov line
in the usual deconfined phase of lattice gauge theory. 

At first glance these plots suggest that
the scalars are driven to the origin as $\beta\to\infty$ (or
equivalently $L\to\infty$ at fixed $\mu$) 
and the quantum continuum theory possesses only the trivial
vacuum state $<\phi>=0$. However, we must be
be careful as the distributions clearly possess long tails extending
out to large eigenvalue. We have examined this issue in more detail and
indeed find good evidence for power law behavior in the tails of the 
distribution -- see figure~\ref{power}. which shows
a plot of
$\log{(P(\lambda))}$ versus $\log{\lambda}$ using data from
the tail of the distribution for both $SU(2)$
and $SU(3)$ at a coupling $\beta=4.0$. 
The exponent extracted from a linear
fit lies in the range of (minus) $2.1-2.3$.
Such an exponent would clearly lead to a normalizable distribution
and hence a convergent partition function\footnote{Actually
since here we use periodic boundary conditions on all fields the
partition function is more properly thought of as a Witten
index.}. Moreover, since the global
$U(1)$ symmetry cannot break in two dimensions,
it is clear that the first moment of this
distribution $<\lambda>$ will vanish. However, a 
value of $p\le 3$ would ensure that
the variance $<\lambda^2>-<\lambda>^2$ is divergent. 
This can be seen explicitly in
our simulations with
figure~\ref{monte}. showing the Monte Carlo evolution of $\lambda^2$
(averaged over the lattice and the number of colors) for the $\beta=4$
$SU(2)$ run. We see large fluctuations occurring at intervals of
order $10,000$ RHMC trajectories. These spikes make it extremely difficult
to determine the expectation value of $<\lambda^2>$ -- the
statistical error does not decrease with the square root of the
number of measurements as
would be expected for a typical Monte Carlo process.
These large spikes in $<\lambda^2>$ are also seen for $\beta=0.5$ and
$\beta=8.0$.

We conjecture that the existence of these tails is related to
the existence of a noncompact classical moduli space. This
motivates us to consider 
the lattice theory in the limit of very large $\beta$. In this limit
we may restrict our attention to fields which are constant over
the lattice and the theory reduces to a zero dimensional matrix
model. This model has been studied previously
with the result that the eigenvalue
distributions again have power law tails
with (negative) exponent $p\sim 3$ 
{\it independent} of the number of colors $N$ \cite{staudacher}.

The results for finite $\beta$ in the two dimensional models look
rather similar.
It is tempting to
interpret this divergence of the variance $<\lambda^2>-<\lambda>^2$
as evidence that
quantum effects will destroy any classical
vacuum with $<\lambda>\ne 0$ - the fluctuations around any
such solution will be so large as to swamp the mean value and
lead to a restoration of the $\lambda\to-\lambda$ symmetry.

\section{Conclusions}
In this paper we have extended our previous numerical study of ${\cal N}=2$
super Yang-Mills theory in two dimensions described in \cite{susysim}. 
The discretization
we have employed was first proposed in \cite{n=2} and is based upon
mapping the twisted continuum theory into a theory of K\"{a}hler-Dirac
fields which may then be discretized preserving gauge invariance
and a single supersymmetry while simultaneously avoiding problems of fermion
doubling. In the previous paper we presented numerical results supporting
the existence of an exact scalar supersymmetry $Q$ in
the case of the $SU(2)$ theory. In this paper
we have derived the action of the remaining twisted supersymmetries on
the component K\"{a}hler-Dirac fields and studied some non-trivial
Ward identities following from those symmetries for
both $SU(2)$ and $SU(3)$. It should be noted that
many of these additional Ward identities are automatically zero
because of gauge invariance and so the additional symmetries supply
rather few additional constraints. 

Our numerical results, carried
out within the phase quenched approximation, support a restoration
of {\it all} supersymmetries in the continuum limit {\it at least for
large continuum dimensionless coupling $\mu=g^2A$}. Such a limit
corresponds to large physical volumes and appears necessary
in order to justify the phase quenched approximation which we have,
of necessity, employed. Furthermore, we have shown that
an additional suppression of
this phase on finite lattices can be achieved by using thermal boundary
conditions. We have also
examined a correlation function of the scalar fields which shows
good evidence of a restoration of rotational invariance at large
distances which is also encouraging. 

These conclusions are
consistent with perturbative arguments based on continuum 
power counting see for example \cite{kaplan,sugino} which argue that
a single exact supersymmetry plus gauge invariance and the
additional global $U(1)$ symmetry prohibit the occurrence of {\it relevant}
SUSY violating operators as the cut-off is removed.

Finally, we present new results concerning the eigenvalue distributions
of the
scalar fields. For $SU(2)$ we see two symmetrically
disposed peaks in
this distribution
which narrow and approach the origin as $\beta\to\infty$. 
In the case of $SU(3)$ three peaks are seen. This pattern appears
to persist at larger $N$ with the distribution showing $N$ peaks.
A more
detailed analysis reveals power law tails to these distributions 
which ensures
that sufficiently high moments of the eigenvalue distributions will diverge.
Our data is consistent with a divergence of the second moment
for both $SU(2)$ and $SU(3)$ at least for large $\beta$. Such a result
is similar to previous results for supersymmetric matrix models
\cite{staudacher}.

Furthermore, we have argued that the divergent variance
would lead to large fluctuations in the scalar fields
which would tend to wash out
any classical vacuum state with $<\phi>\ne 0$. 
Thus the classical moduli space does not survive in the full
quantum theory.

\acknowledgments
This work
was supported in part by DOE grant DE-FG02-85ER40237. Many of the
numerical calculations were carried out using USQCD resources.
The author would
like to thank Joel Giedt and Toby Wiseman for useful discussions.

\vfill
\newpage

\FIGURE{
\epsfig{file=action.eps,width=12cm}\caption{$S_B$ vs $L$ for
$SU(3)$ with $\mu=0.25,5.0$}\label{action}}
\FIGURE{
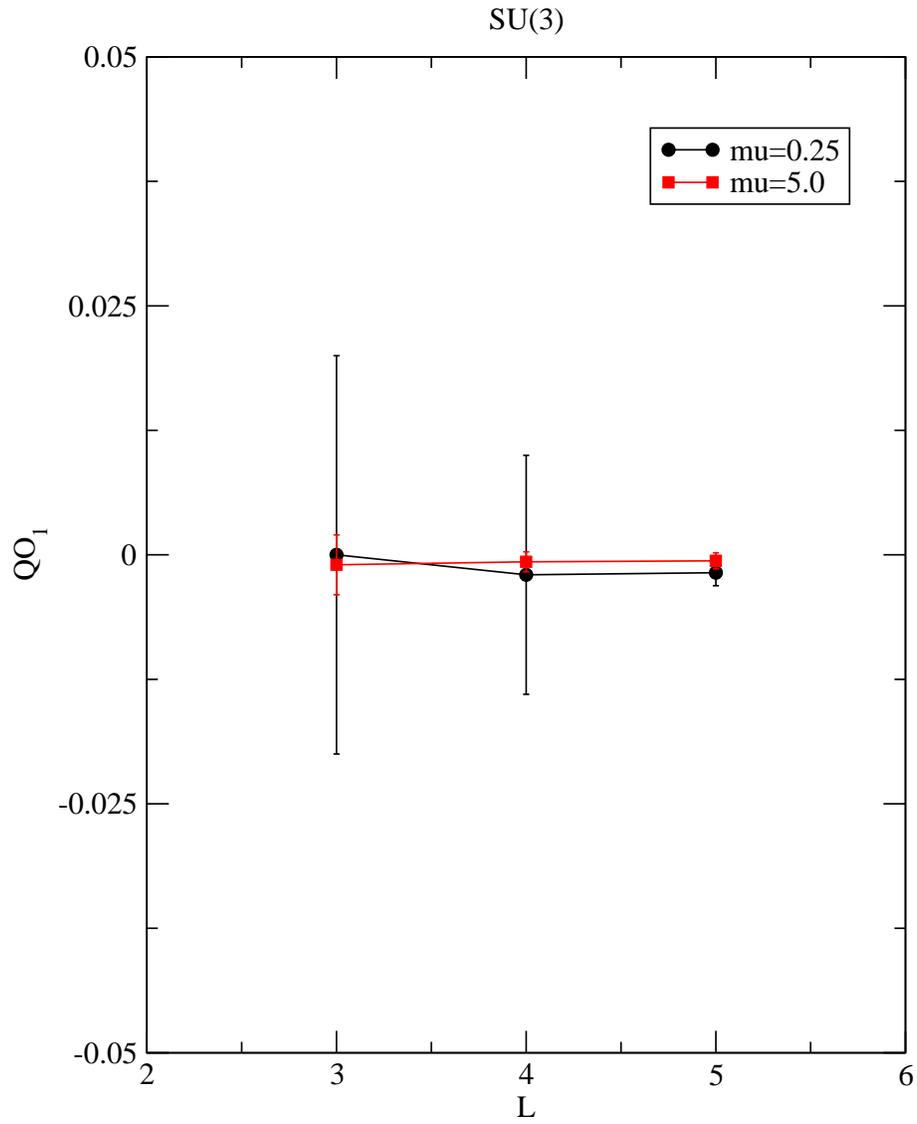
\epsfig{file=o1_all.eps,width=12cm}\caption{$QO_1$ vs $L$ for 
$SU(3)$ with $\mu=0.25,5.0$}\label{o1}}
\FIGURE{
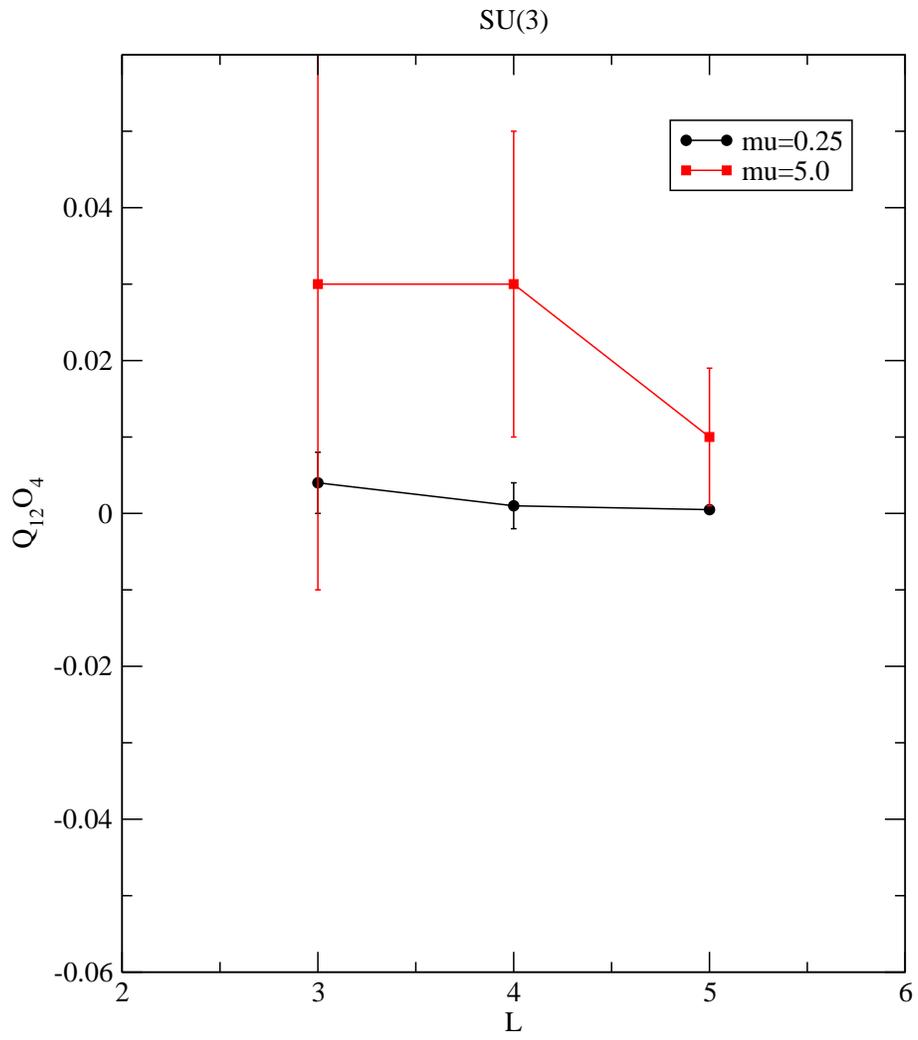
\epsfig{file=o4_all.eps,width=12cm}\caption{$Q_{12}O_4$ vs $L$ for
$SU(3)$ with $\mu=0.25,5.0$}\label{o4}}
\FIGURE{
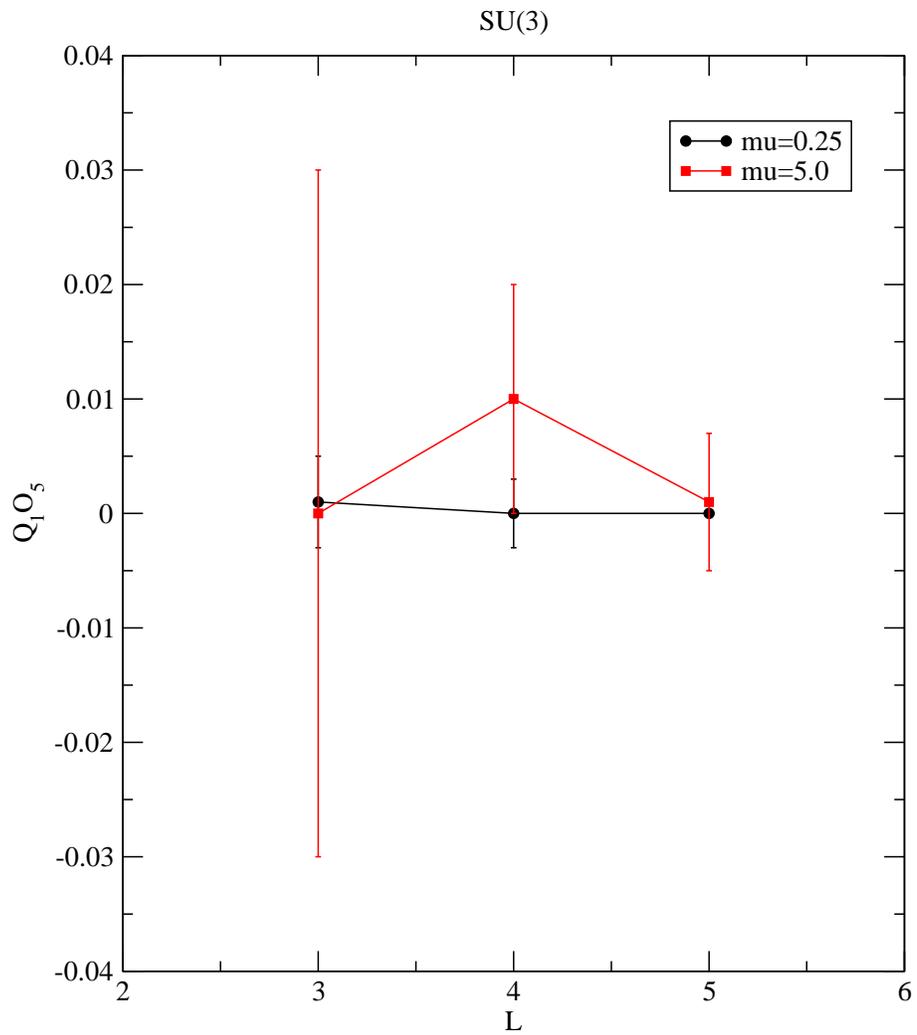
\epsfig{file=o5_all.eps,width=12cm}\caption{$Q_1O_5$ vs $L$ for
$SU(3)$ with $\mu=0.25,5.0$}\label{o5}
}
\FIGURE{
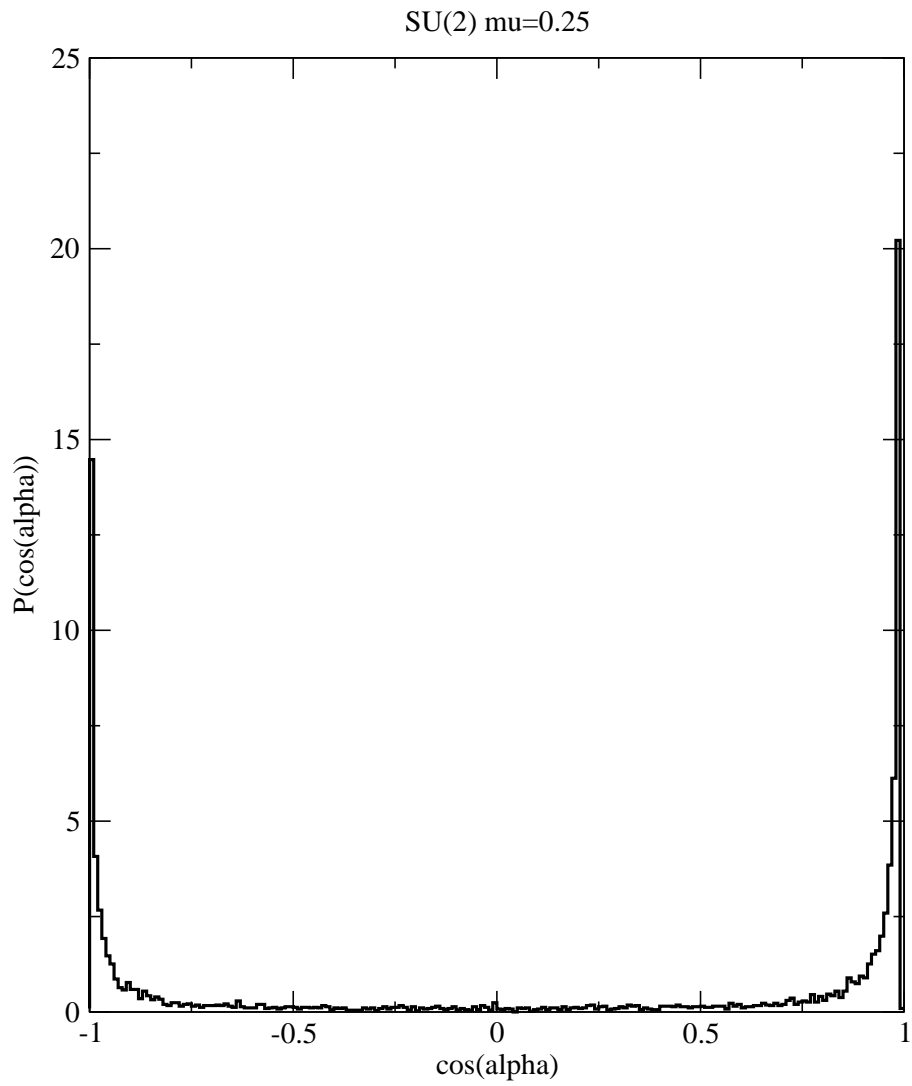
\epsfig{file=l3pbcmu0.25.eps, width=12cm}\caption{$P(\cos{\alpha})$ vs
$\cos{\alpha}$ for $SU(2)$ with $L=3$ $\mu=0.25$}\label{pbcmu0.25}
}
\FIGURE{
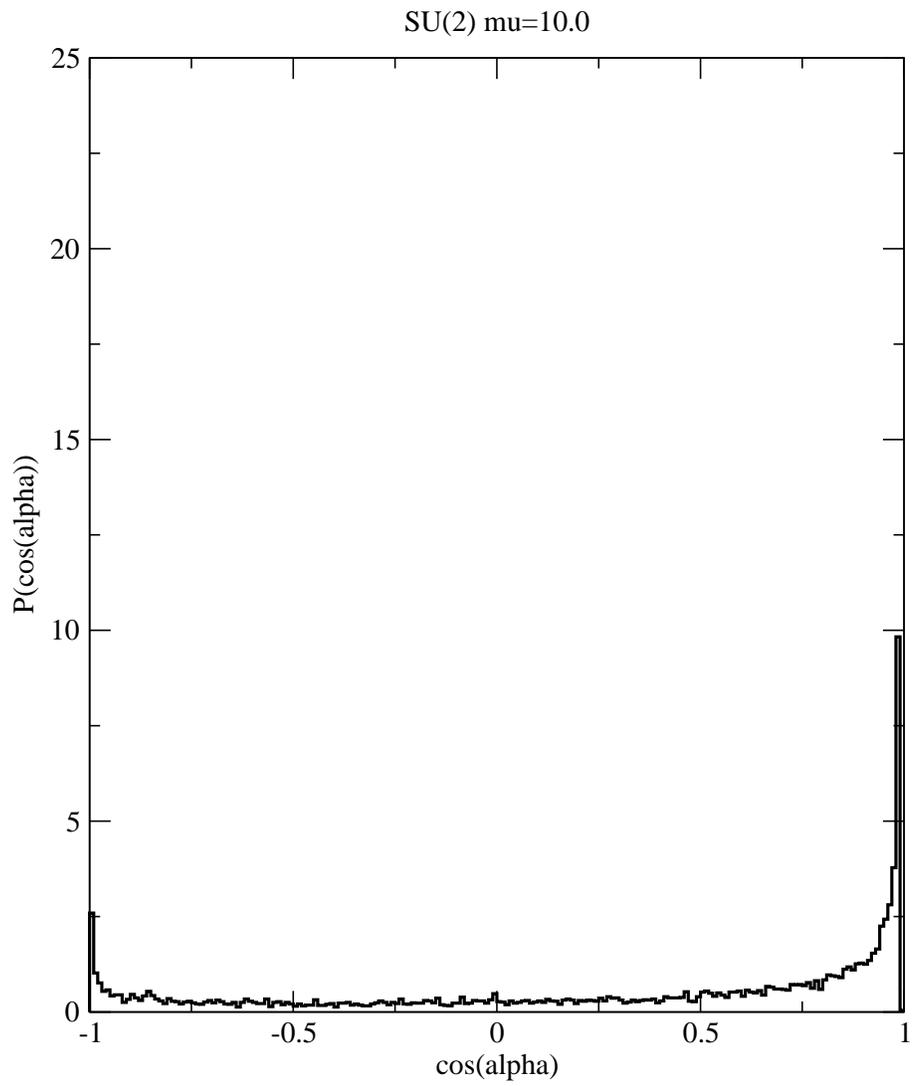
\epsfig{file=l3pbcmu10.0.eps, width=12cm}\caption{$P(\cos{\alpha})$ vs
$\cos{\alpha}$ for $SU(2)$ with $L=3$ $\mu=10.0$}\label{pbcmu10.0}
}
\FIGURE{
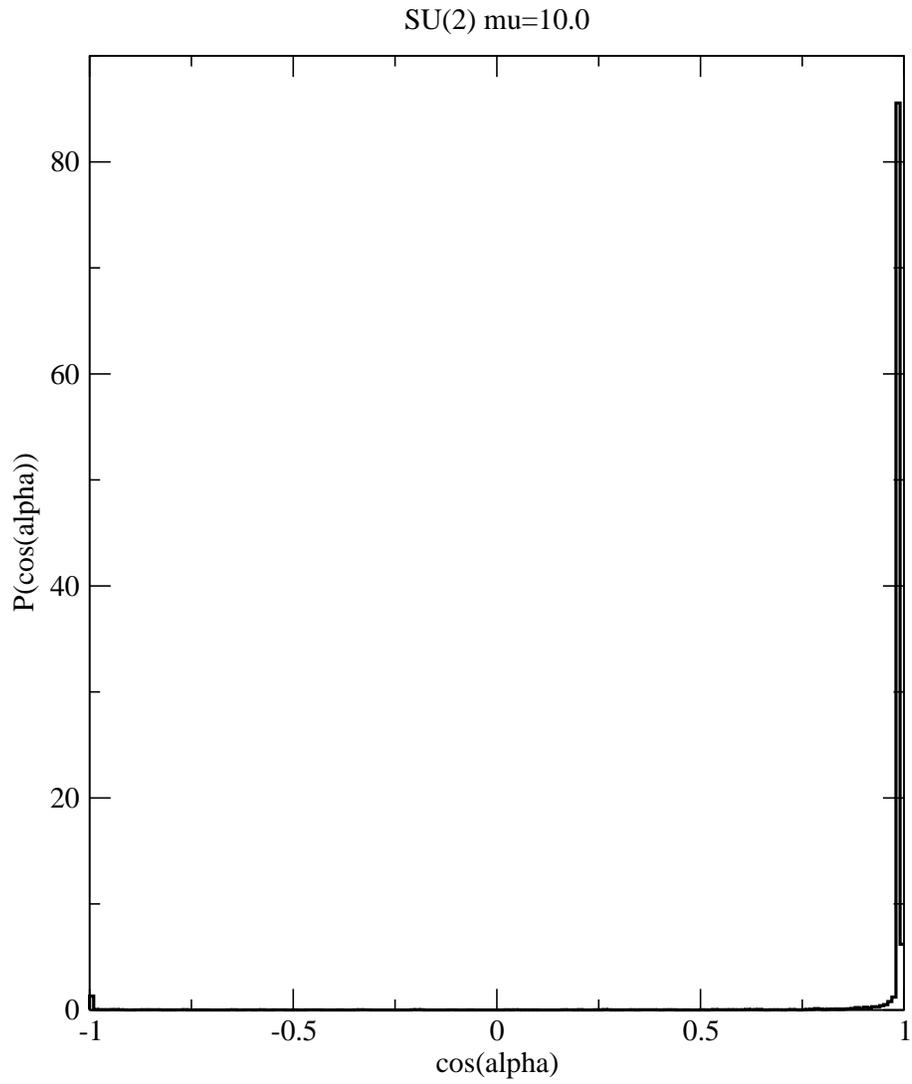
\epsfig{file=apbcmu10.0.eps, width=12cm}\caption{$P(\cos{\alpha})$ vs
$\cos{\alpha}$ for thermal $SU(2)$ with $L=3$ $\mu=10.0$}\label{apbcmu10.0}
}
\FIGURE{
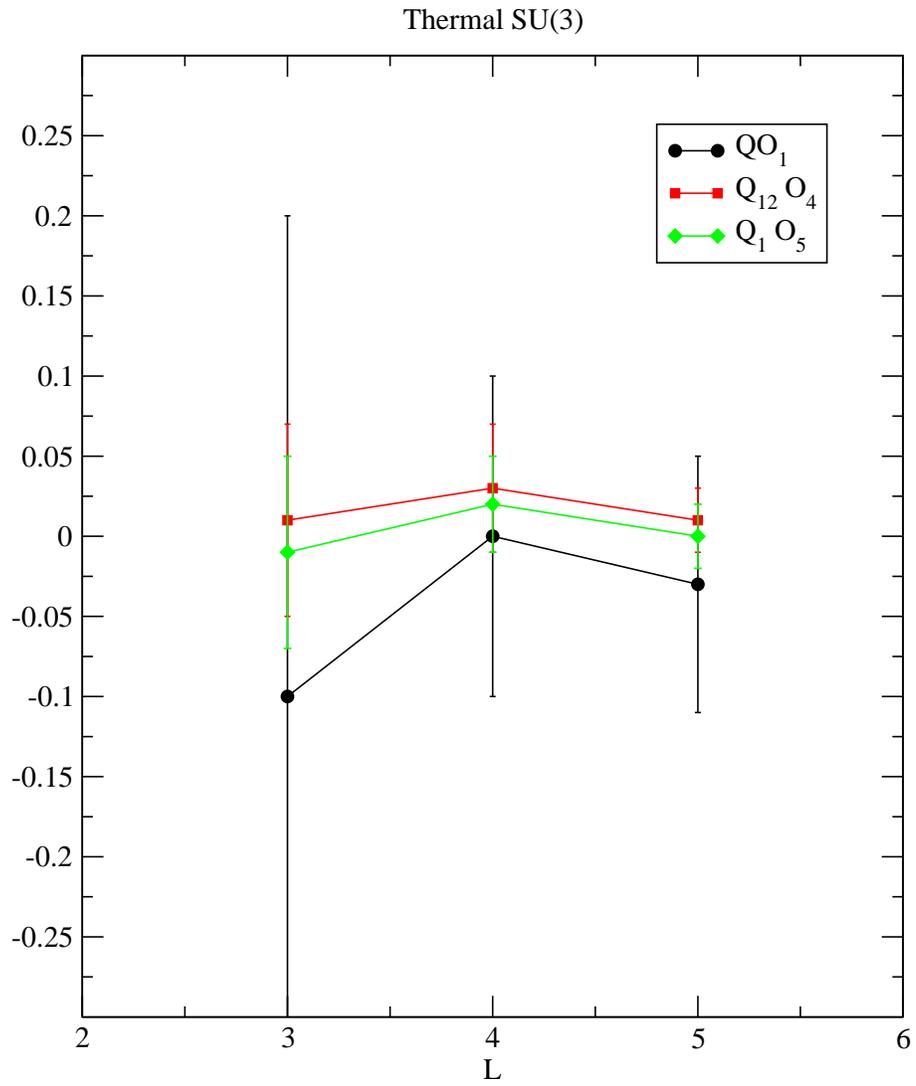
\epsfig{file=apbcward.eps,width=12cm}\caption{$QO_1$, $Q_{12}O_4$ and $Q_1O_5$
vs $L$ for thermal $SU(3)$ at $\mu=5.0$}\label{apbcward}
}
\FIGURE{
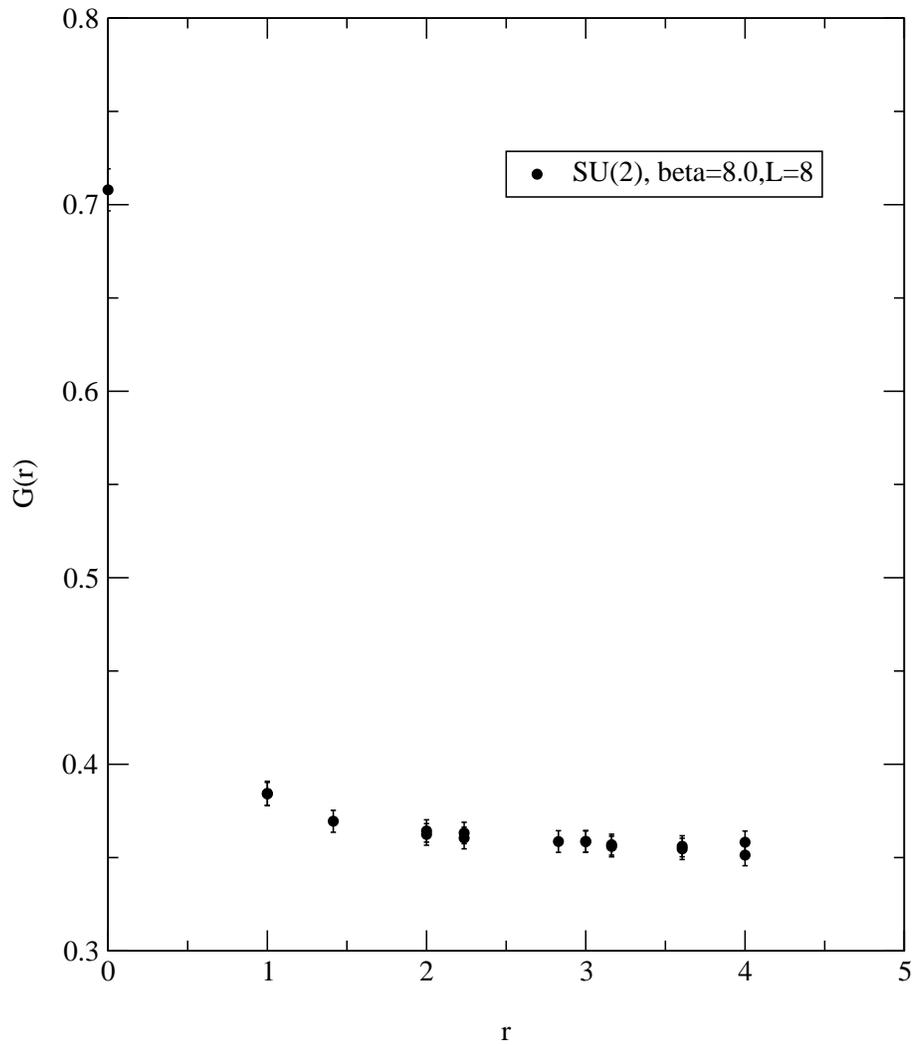
\epsfig{file=corrsu2.eps,width=12cm}\caption{$G(r)$ vs $r$ for $SU(2)$ and
$\beta=8.0$ with $L=8$}\label{corrsu2}
}
\FIGURE{
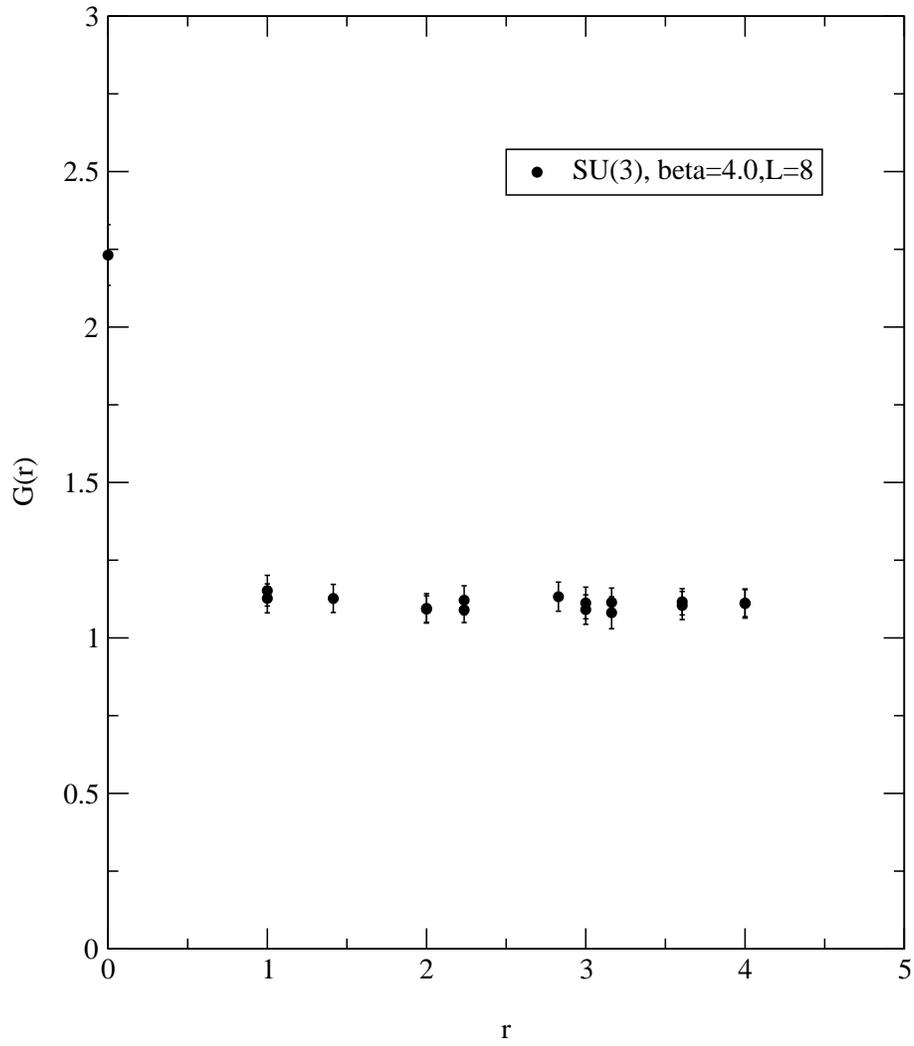
\epsfig{file=corrsu3.eps,width=12cm}\caption{$G(r)$ vs $r$ for $SU(3)$
and $\beta=8.0$ with $L=8$}\label{corrsu3}
}
\FIGURE{
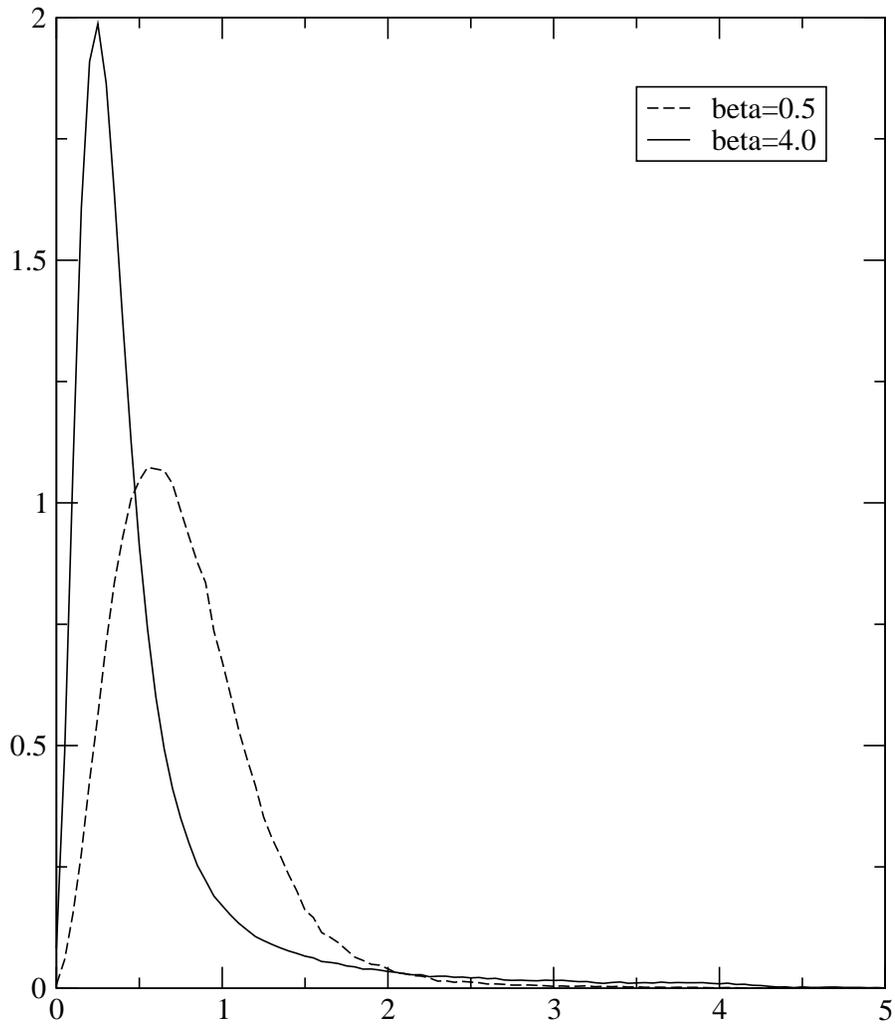
\epsfig{file=su2.eps,width=12cm}\caption{Eigenvalue distribution for $SU(2)$
and $L=3$}\label{su2}
}
\FIGURE{
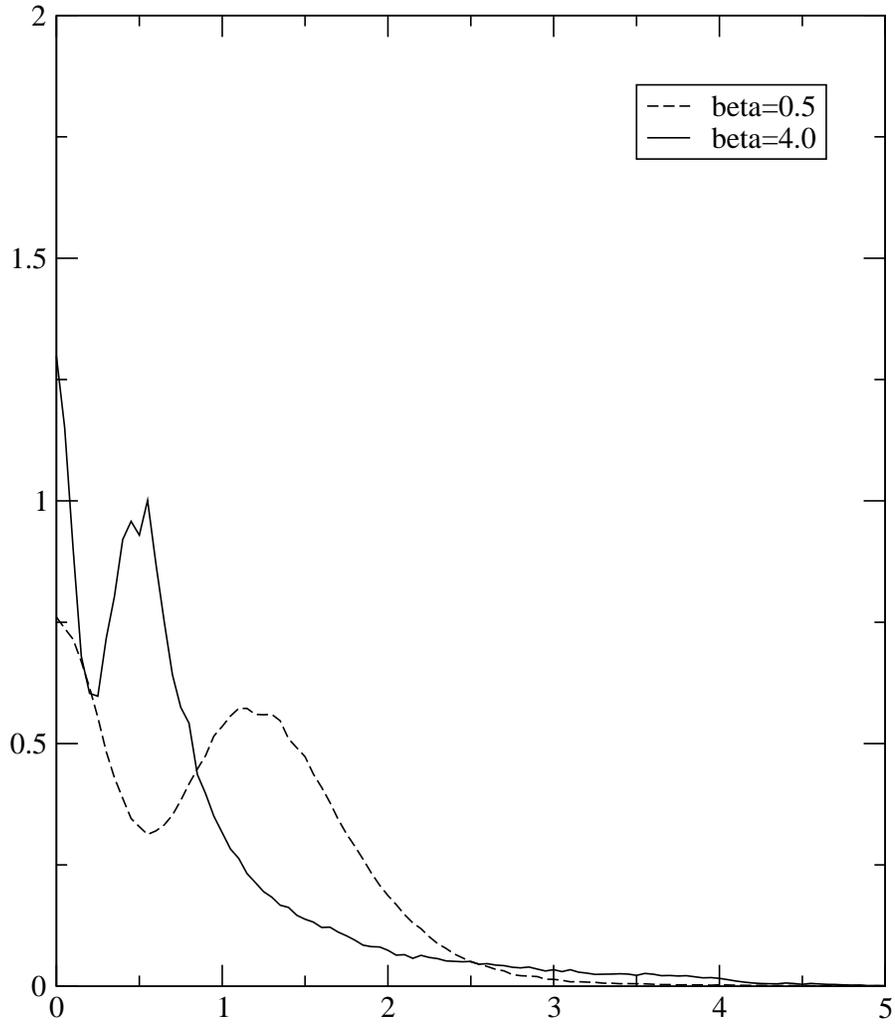
\epsfig{file=su3.eps,width=12cm}\caption{Eigenvalue distribution for $SU(3)$
and $L=3$}\label{su3}
}
\FIGURE{
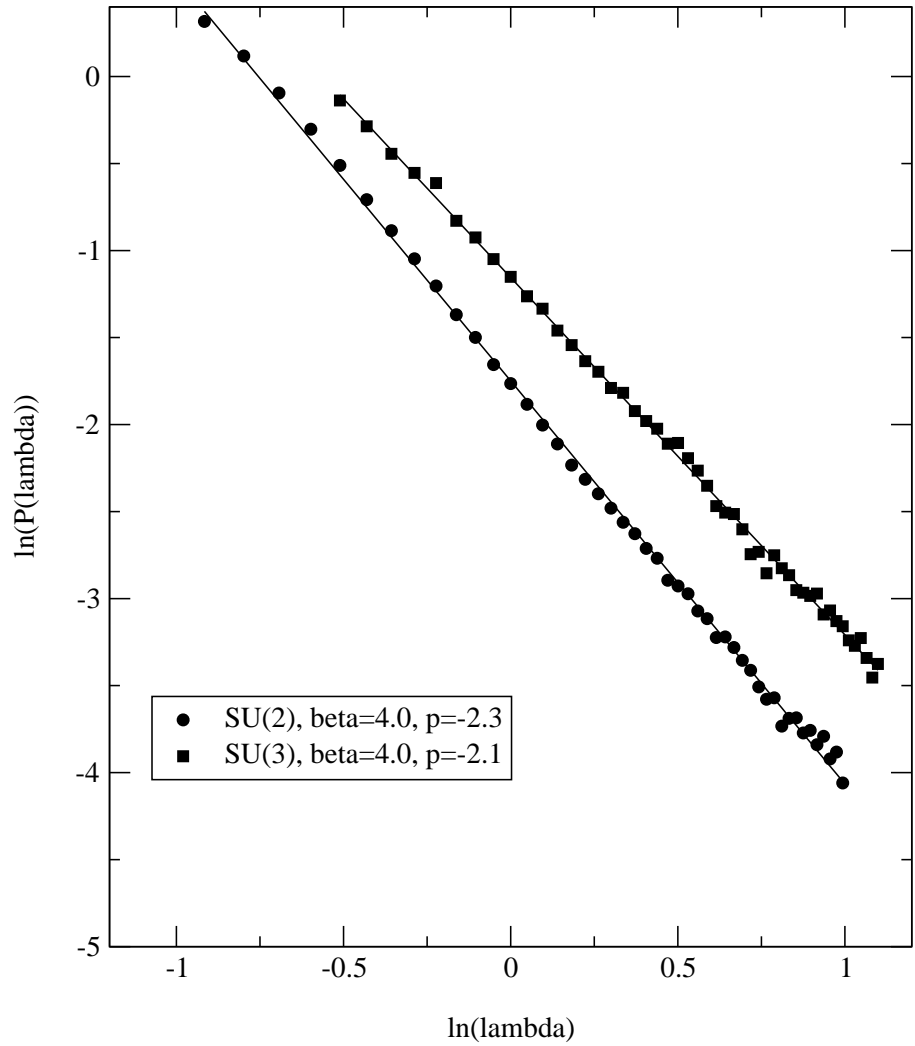
\epsfig{file=power.eps,width=12cm}\caption{$\log{(P(\lambda)}$ vs
$\log{\lambda}$ for $L=3$ and both $SU(2)$ and $SU(3)$}\label{power}}

\FIGURE{
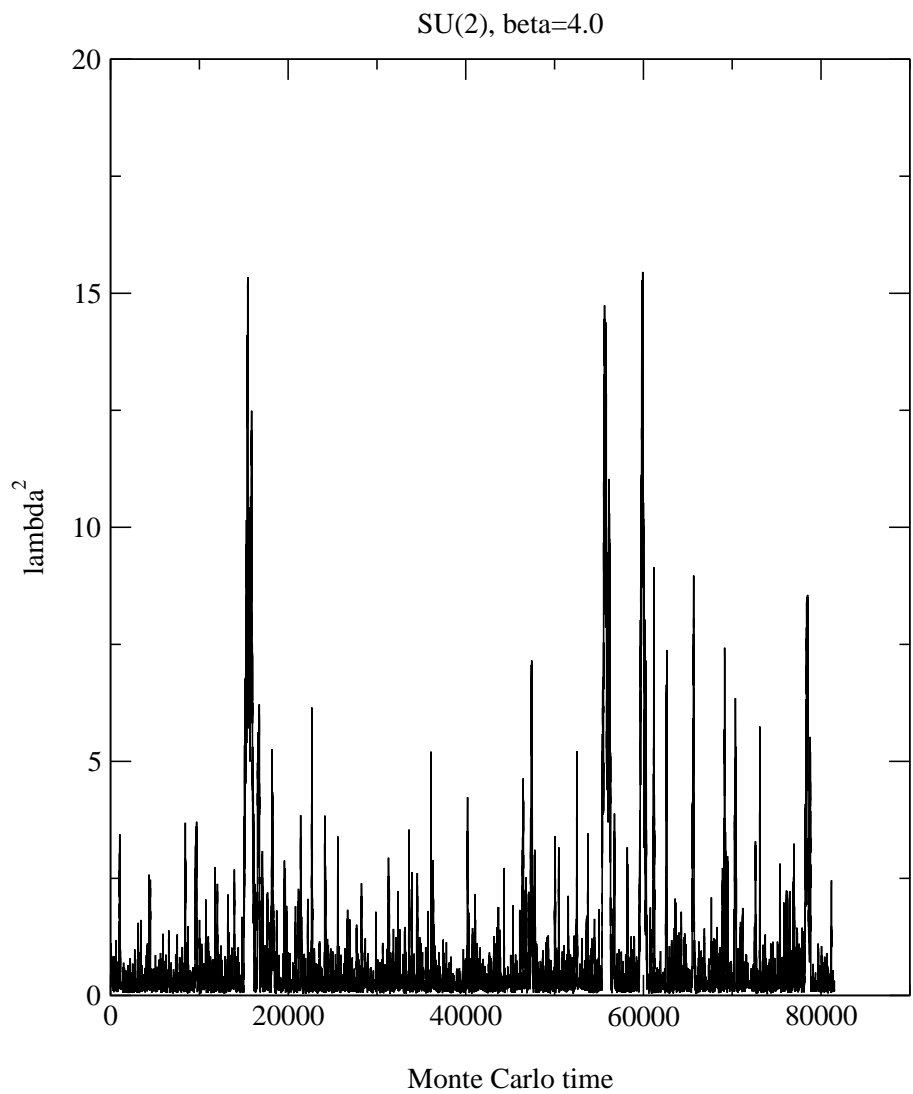
\epsfig{file=monte.eps,width=12cm}\caption{Monte Carlo history
for $L=3$, $\beta=4.0$ and $SU(2)$}\label{monte}}

\end{document}